\documentclass[%
reprint,    
showpacs,
showkeys,
amsmath,amssymb,
aip,
jcp,
letter
]{revtex4-2}

\usepackage{graphicx} 
\usepackage{dcolumn} 
\usepackage{bm} 
\usepackage[page]{appendix}
\usepackage{amsmath}
\usepackage{float}
\usepackage{xcolor}
\usepackage{scalerel}
\usepackage{subfigure}
\usepackage{color}
\usepackage{etoolbox}
\usepackage[hidelinks]{hyperref} 
\hypersetup{
    colorlinks=true,
    linkcolor=blue,
    citecolor=blue,
    filecolor=magenta,      
    urlcolor=cyan,
}

\begin{document}

\title{Thermodynamic Descriptors from Molecular Dynamics as Machine Learning Features for Extrapolable Property Prediction}

\author{Nuria H. Espejo$^\dag$}
\affiliation{Department of Physical-Chemistry, Universidad Complutense de Madrid, Av. Complutense s/n, Madrid 28040, Spain}
\affiliation{Data Science, Bayer AG, Alfred-Nobel-Straße 50, 40789 Monheim am Rhein, Germany}

\author{Pablo Llombart$^\dag$}
\affiliation{Department of Physical-Chemistry, Universidad Complutense de Madrid, Av. Complutense s/n, Madrid 28040, Spain}

\author{Andrés González de Castilla}
\affiliation{Crop Protection Innovation, Bayer AG, Kaiser-Wilhelm-Alle 1, 51373 Leverkusen, Germany}

\author{Jorge Ramirez}
\email{jorge.ramirez@upm.es}
\affiliation{Department of Chemical Engineering, Universidad Polit\'ecnica de Madrid, C/ Jose Gutierrez Abascal 2, Madrid 28006, Spain}
\author{Jorge R. Espinosa}
\email{jorgerene@ucm.es}
\affiliation{Department of Physical-Chemistry, Universidad Complutense de Madrid, Av. Complutense s/n, Madrid 28040, Spain}
\affiliation{Yusuf Hamied Department of Chemistry, University of Cambridge, Lensfield Road, Cambridge CB2 1EW, UK \\ $\dag$ These authors contributed equally to this work.}

\author{Adiran Garaizar}
\email{adiran.garaizarsuarez@bayer.com}
\affiliation{Data Science, Bayer AG, Alfred-Nobel-Straße 50, 40789 Monheim am Rhein, Germany}

\date{\today}

\begin{abstract}
The limited extrapolative power of structure-based machine learning (ML) models is a critical bottleneck in chemical discovery, particularly for industrial R\&D, where navigating uncharted chemical space to find next-generation materials or drugs is paramount. These models, reliant on structural descriptors or graph neural networks (GNNs), often fail when predicting properties for molecules with novel chemotypes. Here, we introduce a physics-augmented ML framework that overcomes this limitation. Our approach replaces conventional structural inputs with thermodynamic properties such as cohesive energy, heat of vaporization, and density, derived directly from molecular dynamics (MD) simulations. While performing comparably to structure-based models on known organic compounds, our method uniquely maintains low error when extrapolating to dissimilar chemical spaces. Crucially, it accurately predicts boiling points for entire chemical classes absent from the training set, including inorganic compounds, salts, and molecules with elements such as Si, B, and Te. By learning from the intermolecular forces that govern phase transitions, our framework provides a more fundamental and generalizable strategy for molecular property prediction, enabling chemical exploration beyond established structural domains.

\end{abstract}


\maketitle

\section{\label{sec:intro}Introduction}
{The accurate prediction of macroscopic material properties from molecular structure is a foundational goal of computational chemistry, underpinning fields from drug discovery to materials science~\cite{schneider2018automating,mak2024artificial}. The utility, safety, and commercial viability of new compounds are dictated by a suite of physicochemical properties, including its solubility, viscosity, and phase behavior~\cite{van2003admet,reid1959properties}. Among this diverse landscape of properties, those governing phase transitions are particularly fundamental. The normal boiling point (nBP), for instance, serves as a key benchmark for testing the generalization of any predictive framework~\cite{egolf1994prediction}. This property is of immense practical importance, with broad relevance in pharmaceutical manufacturing, chemical engineering, sustainability, and safety assessment~\cite{seader2016separation,jessop2011searching,prat2016sanofi,poling2001properties}. However, experimental determination, while well-established for many common substances, can be a low-throughput endeavor, with methodologies and costs that vary significantly based on the substance in question~\cite{bell2018pure,kim2025pubchem,sandler2017chemical,poling2001properties}. This challenge is particularly pronounced for difficult-to-measure or model compounds, such as active pharmaceutical ingredients (APIs)~\cite{wu2023continuous}, difficult-to-isolate systems including impurities~\cite{castro2025understanding,finotti2026impurities}, salts and ionic liquids~\cite{brennecke2001ionic}. This gap between the industrial demand for reliable data and the limitations of current predictive tools has been identified as a key challenge for the future of applied thermodynamics, underscoring the urgent need for more robust and transferable models capable of navigating the novel chemical spaces essential for industrial innovation and intellectual property.~\cite{de2022view}.}

{A rich spectrum of methods has been developed to address nBP and vapor-pressure prediction. Classical approaches, such as corresponding-states methods~\cite{lee1975generalized,ambrose1989vapour}, exploit reduced-property correlations based on critical data, but their scope is limited when such data is unavailable, or the target molecule's chemotype diverges significantly~\cite{bruce2001properties}. Group-contribution (GC) schemes~\cite{edwards1981estimation,tu1994group,constantinou1994new,marrero1999estimation,asher2006vapor,pankow2008simpol,moller2008estimation,ceriani2013prediction,rezakazemi2013development,wang2015predicting}, like the well-known Joback~\cite{joback1987estimation} or Nannoolal~\cite{nannoolal2004estimation} methods, decompose molecules into predefined structural fragments. While rapid, and highly effective for molecules that can be segmented into their parameterized groups, GC methods are structurally blind to compounds built from elements or motifs not encoded in their parameterized fragment libraries~\cite{alibakhshi2022dependence}. In a similar vein, equation-of-state (EoS) models like Peng–Robinson or SAFT~\cite{peng1976new,chapman1989saft,gross2001perturbed} can offer physically motivated predictions, but only when reliable component-specific parameters are known or sufficient experimental data is available for their parameterization.}

{The last two decades, however, have seen a revolution driven by the modern data-driven paradigm of Quantitative Structure-Property Relationship (QSPR) modeling~\cite{katritzky1995qspr}. This approach uses pre-calculated molecular descriptors to train a statistical model, from early efforts with multiple linear regression~\cite{katritzky2007rapid,hilal2003prediction,dai2013prediction,mirshahvalad2019neural,fissa2019qspr} to more advanced algorithms like Random Forest~\cite{kim2024integrating} or Gradient Boosting~\cite{zhang2014qspr}. While powerful, the predictive capability of all these models is fundamentally limited by the expressiveness and transferability of manually engineered descriptors~\cite{koam2025machine}.}

{A more recent breakthrough has been the evolution to deep learning, particularly Graph Neural Networks (GNNs), which they learn property-relevant representations directly from the molecular graph, bypassing the need for manual descriptor engineering~\cite{duvenaud2015convolutional,gilmer2017neural,xiong2019pushing}. This paradigm has been successfully applied to a wide array of thermophysical properties~\cite{rittig2023gibbs, damay2021predicting, specht2024hanna,ahmad2023attention,winter2025understanding, felton2024ml, habicht2023predicting}, and it has become a dominant strategy for nBP prediction~\cite{qu2022graph, sangala2025graph,lin2024advancing, santana2024puffin}. This approach has been notably successful, with state-of-the-art frameworks like GRAPPA~\cite{hoffmann2025grappa}, achieving exceptional accuracy. Trained on tens of thousands of experimental data points, GNNs routinely outperform classical schemes within their training domain~\cite{hoffmann2025grappa}. Indeed, recent work has shown that specialized GNNs can achieve high accuracy even for notoriously complex systems like ionic liquids, provided they are trained on vast, domain-specific datasets \cite{rittig2023graph}.  However, because they learn statistical associations from molecular topology without direct access to the thermodynamic state of the liquid, their performance is tightly coupled to the training data's diversity~\cite{yang2019analyzing}. When applied to molecules that are structurally and physicochemically dissimilar from the training distribution, their extrapolative behavior becomes uncertain~\cite{tetko2020limitations}. This effectively excludes many classes of novel, drug-like structures from their reliable applicability domain~\cite{janet2019designing,wu2018moleculenet}. This limitation is particularly significant in industrial settings, where the primary goal is to generate intellectual property by exploring precisely these novel chemical domains.}

{The limitations of purely structural models motivate a direct solution: re-introducing the missing physics by using physics-based descriptors. This approach moves beyond the pre-defined structural descriptors---ranging from 2D fragments to 3D conformational properties---common in traditional QSPR, and instead computes features derived directly from first-principles simulations. One category of these descriptors is derived from quantum mechanics (QM), which provides a highly accurate description of a single molecule's electronic structure~\cite{gugler2023machine,bryce2011physics,li2024quantum}. However, while physically rigorous, these descriptors inherently describe a molecule in isolation, and thus neglect the crucial intermolecular interactions and ensemble effects that govern a condensed-phase property like the boiling point, a subtlety that can lead to misleading conclusions if not interpreted carefully~\cite{higginbotham2024predicting}. A more sophisticated approach, COSMO-RS, attempts to bridge this gap~\cite{klamt2000cosmo}. It uses QM to predict thermodynamic properties for a wide range of chemicals by placing the molecule in a virtual conductor, thereby modeling the ensemble effect via a continuum~\cite{zhang2023estimation,song2025prediction,dupeux2022cosmo}. Yet, as a continuum-based model, COSMO-RS averages intermolecular interactions, leading to documented quantitative failures for systems governed by strong, specific forces, such as those in ionic liquids~\cite{kurnia2014evaluation,paduszynski2017overview}. A more direct approach is to use molecular dynamics (MD) simulations to explicitly model the molecular ensemble~\cite{frenkel2002understanding,koeddermann2007molecular,smith2014molecular}. Indeed, computational studies have long leveraged MD to investigate phase transitions and related thermodynamic properties, from calculating vaporization enthalpies of complex fluids like ionic liquids~\cite{zaitsau2019isolating} to characterizing liquid-vapor interfaces~\cite{harris1992liquid}, and determining phase equilibria from rigorous but computationally demanding free energy calculations~\cite{luo2022unified,emamian2022performance}. However, these computational efforts have traditionally focused on either highly demanding, direct coexistence simulations to compute the nBP in the liquid-vapor equilibrium from scratch~\cite{panagiotopoulos1987direct}, or on using simulation data to develop more accurate force fields~\cite{nitzke2025ms2}. The computational expense of such simulations, however, has remained a significant barrier to their routine use for generating descriptors at scale, creating a need for more efficient methodologies. The challenge, therefore, is to create a hybrid framework that harnesses the predictive efficiency of data-driven models while being firmly anchored in the physical principles that govern molecular behavior. This goal of anchoring data-driven models in physical principles is central to the emerging paradigm of Physics-Informed Machine Learning (PIML)~\cite{karniadakis2021physics, meng2025when, kashinath2021physics, raissi2019physics, zhu2023physics, sosso2019harnessing, wu2018physics,jirasek2020hybridizing}.}

{In this work, we implement a physics-augmented strategy specifically designed for scalable and extrapolative property prediction. We hypothesize that thermodynamic features derived from computationally tractable MD simulations can unlock robust generalization where purely structural models fail. We run short, all-atom simulations to compute descriptors such as the cohesive energy, and use these quantities as features to train a CatBoost~\cite{prokhorenkova2018catboost} regression model. We show that this enables reliable extrapolation beyond the traditional domain of QSPR, successfully predicting boiling points for chemical classes where structural models are often fragile or inapplicable, including inorganic compounds, complex charged systems like salts, and molecules containing uncommon elements such as tellurium and boron.}

\begin{figure*}[htb!]
    \centering
	\includegraphics[width=\textwidth]{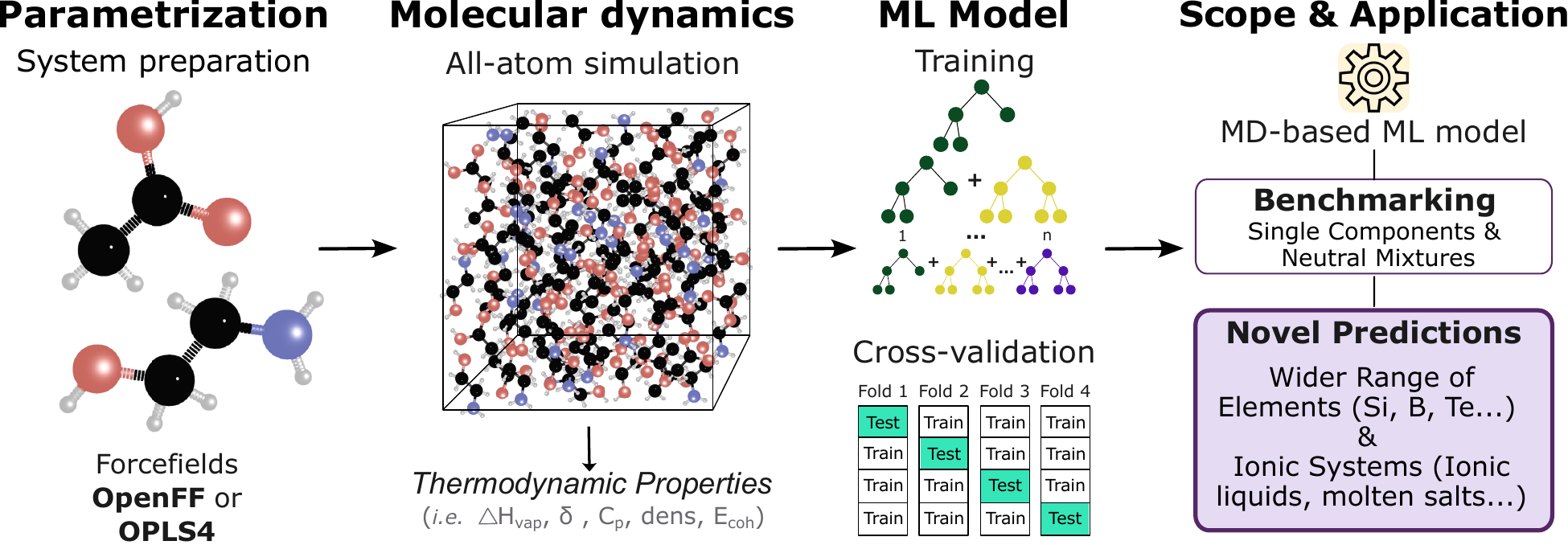}
	\caption{Schematic of the physics-augmented workflow for normal boiling point prediction. Molecular structures are built from SMILES strings and subjected to short all-atom molecular dynamics simulations using the OPLS and OpenFF forcefields. Thermodynamics properties, including cohesive energy and heat of vaporization, are extracted from the simulations and used as physics-augmented  descriptors to train a CatBoost regression model for the final boiling point prediction.}
	\label{fig: Fig1}
\end{figure*}

\section{Materials and Methods}\label{sec: Materials and Methods}
\subsection{Training and Test Sets}

{The initial training dataset was sourced from the curated compilation in Ref.\cite{kim2024integrating}, which provides boiling point data for approximately 1,748 organic compounds (primarily hydrocarbons, alcohols, and amines) originally extracted from the CAS database~\cite{cas_common_chemistry}. To ensure the highest data quality, all experimental values were subsequently cross-verified against multiple authoritative databases: the NIST Chemistry WebBook\cite{linstrom2001nist}, PubChem~\cite{kim2025pubchem}, and ChemSpider~\cite{pence2010chemspider}. Compounds were excluded from our final dataset if they exhibited discrepancies greater than 10~K across sources, lacked corroborating entries, or were flagged with uncertainty markers. To ensure the physical validity of the simulations, a molecular weight threshold of 225 g/mol was applied. This step was necessary to exclude high-molecular-weight compounds that would likely be solid or crystalline. In either case, the simulation would not represent a stable liquid-phase ensemble, invalidating the basis for our descriptor calculation. This filtering removed approximately 500 compounds, yielding a final training set of 1,280 entries with high-confidence experimental values. This collection forms the basis of our Training Set (available in section \ref{sec:Data Availability}).}

{To rigorously evaluate model performance, two distinct test sets were curated, each designed to probe a different aspect of predictive capability. The first, a benchmark set for extrapolation, comprises active pharmaceutical ingredients (APIs) sourced from DrugBank~\cite{wishart2024drugbank}. These molecules were selected for their high structural complexity, often featuring more intricate scaffolds and a greater diversity of functional groups compared to the simpler, smaller molecules constituting the training set. These molecules, while structurally complex and diverse, are composed of standard organic elements. This composition ensures they fall within the theoretical applicability domain of modern GNNs, enabling a direct, head-to-head comparison of our model's extrapolative performance against state-of-the-art methods on challenging, real-world structures.}

{The second, an out-of-domain challenge set, was curated to challenge the model's generalization capabilities on chemical classes that lie far outside the applicability domain of traditional QSPR. It comprises chemotypes---including ionic liquids, salts, and molecules containing non-standard elements such as B, Si, and Te---for which many standard models are fundamentally inapplicable as they cannot generate a prediction for structures containing unparameterized elements or fragments. As a direct benchmark is therefore often impossible, predictions for this set are reported in order to characterize the model's behavior and define the boundaries of its applicability in these challenging chemical spaces.}

\subsection{Molecular Dynamics Simulations}
{To generate the thermodynamic descriptors for our model, we performed atomistic molecular dynamics (MD) simulations for all compounds. The following protocol constitutes the workflow necessary to generate features for each compound. To assess the robustness and transferability of the descriptors, we established two parallel and independent simulation workflows, each employing a different state-of-the-art force field.

The OpenFF workflow used the OpenFF-2.0.0 'Parsley' force field~\cite{wagner2021openforcefield}, with simulations run in the GROMACS 2025.2 package~\cite{van2005gromacs}. Separately, the OPLS4 workflow was run using the OPLS4 force field~\cite{lu2021opls4} as implemented in Schrödinger’s Materials Science Suite.

For each compound and for each of the two force fields, three independent NPT simulations of 20 ns were performed at 300, 400 and 500 K, all at 1 atm with a 2 fs integration time step. Simulation boxes contained approximately 15,000 atoms. Thermodynamic properties were averaged over the final 5 ns of each trajectory. From these trajectories, a set of thermodynamic descriptors was computed, including the cohesive energy ($E_{\text{coh}}$),  defined as the negative of the average intermolecular potential energy ($E_{\text{coh}} \approx -\langle U_{\text{inter}} \rangle$), heat of vaporization ($\Delta H_{\text{vap}} \approx E_{\text{coh}} + RT$), density ($\rho$), the Hildebrand solubility parameter ($\delta = \sqrt{(\Delta H_{\text{vap}} - RT)/V_m}$), and the isobaric specific heat capacity ($C_P$). The latter was computed from enthalpy fluctuations via the relation $C_P = (\langle H^2 \rangle - \langle H \rangle^2) / (k_B T^2)$ \cite{allen2017computer}.}

\subsubsection{OpenFF Simulations}

{Molecular topologies were generated from SMILES strings using RDKit~\cite{landrum2013rdkit} and the OpenFF Toolkit, with Gasteiger~\cite{gasteiger1980iterative} partial charges. Initial simulation boxes were built with Packmol~\cite{martinez2009packmol}. Systems were then energy-minimized and equilibrated for 2 ns under NPT conditions in GROMACS 2025.2~\cite{van2005gromacs}.

Production runs were conducted using the leap-frog (md) integrator with a 2 fs time step. Long-range electrostatics were handled via the Particle Mesh Ewald (PME) method~\cite{essmann1995smooth} with a short-range cutoff of 0.9 nm. Van der Waals interactions were treated with a cutoff at 0.9 nm, using a force-switching function that started at 0.8 nm to smoothly bring the forces to zero. Long-range dispersion corrections for energy and pressure were applied. The V-rescale thermostat~\cite{bussi2007canonical} (1.0 ps relaxation time, isotropic) and the C-rescale barostat (2.0 ps relaxation time) were used to maintain temperature and pressure, respectively. All bonds were constrained using the LINCS algorithm~\cite{hess1997lincs}.

To calculate $E_{\text{coh}}$ in this workflow, the system was divided into two equal molecular subgroups. The total intermolecular potential energy between these subgroups (short-range Coulomb + van der Waals) was summed, and the resulting value, $\langle U_{\text{inter}} \rangle$, was inverted in sign and normalized by the number of molecules to yield the final cohesive energy.}

\subsubsection{OPLS4 Simulations}
{Systems were prepared using the Desmond System Builder in Maestro (Schrödinger Materials Science Suite). After energy minimization and multi-stage equilibration, production NPT runs of 20 ns were conducted using the Nosé–Hoover chain thermostat~\cite{evans1985nose} (1 ps relaxation) and Martyna–Tobias–Klein barostat~\cite{evans1985nose} (2 ps, isotropic). The RESPA integrator was employed with bonded/near/far time steps of 2.0/2.0/6.0~fs. Short-range Coulomb interactions were truncated at 0.9 nm. For this workflow, the cohesive energy ($E_{\text{coh}}$) was computed as the negative of the average total intermolecular potential energy, extracted directly from Desmond's simulation output.}


\subsection{Machine Learning Model} {The set of MD-derived descriptors was chosen based on a balance between their direct physical relevance to vaporization and their computational accessibility. Properties like cohesive energy and heat of vaporization directly quantify the intermolecular forces that must be overcome during boiling. Density reflects the packing of molecules in the liquid phase, while the solubility parameter and the isobaric heat capacity provide composite measures of the system's cohesive and thermal properties. Crucially, all these descriptors can be efficiently extracted from standard NPT simulations, providing a rich, physically-grounded feature set for the machine learning model without resorting to more computationally intensive protocols.}

For each force field, three CatBoost \cite{prokhorenkova2018catboost} gradient-boosted regression models were trained:
\begin{enumerate}
    \item \textbf{MD-only model:} Trained exclusively on MD-derived thermodynamic descriptors ($E_{\text{coh}}$, $\Delta H_{\text{vap}}$, $\rho$, $\delta$, $C_P$) extracted at 300, 400 and 500~K. 
    \item \textbf{Chemoinformatics-only models:} Trained on a large set of chemoinformatics descriptors computed from SMILES strings using RDKit \cite{landrum2013rdkit}. This feature set is composed of three main groups: (a) MACCS keys (166 bits), (b) Morgan circular fingerprints (radius 2, 2048 bits), and (c) a wide range of other 2D physicochemical descriptors, including properties like molecular weight, TPSA, hydrogen bond counts, and many others.
    \item \textbf{Hybrid models:} Combined both MD-derived thermodynamic features and structural descriptors into a unified feature set.
\end{enumerate}

Hyperparameter optimization was performed via Bayesian search using Optuna \cite{akiba2019optuna} with 50~trials. The training set (1280 compounds) was split using stratified 4-fold outer cross-validation scheme. The groups were generated by clustering the 1,280 training compounds based on structural similarity, using a sphere exclusion algorithm with Morgan fingerprints and a Tanimoto similarity threshold of 0.65. This procedure ensures that structurally similar molecules are kept within the same fold, forcing the model to be validated on chemical scaffolds that are distinct from those seen during training in each split. For each outer fold, a 3-fold inner cross-validation loop was used for hyperparameter optimization, which was conducted via a 50-trial Bayesian search using Optuna \cite{akiba2019optuna}. Key optimized parameters included the learning rate, tree depth, L2 regularization, and the number of boosting iterations (up to 1000, with early stopping). Feature selection used CatBoost's native importance scores. Our model performance was quantified using the mean absolute error (MAE), root mean squared error (RMSE), and coefficient of determination ($R^2$). Statistical significance was assessed via paired Wilcoxon signed-rank tests~\cite{woolson2007wilcoxon}.

\section{Results}\label{sec: Results}
\subsection{Simulation-derived cohesive energies correlate with experimental boiling points across multiple force fields and temperatures}\label{subsec: Result1}
\begin{figure*}[htb!]
    \centering
	\includegraphics[width=0.8\textwidth]{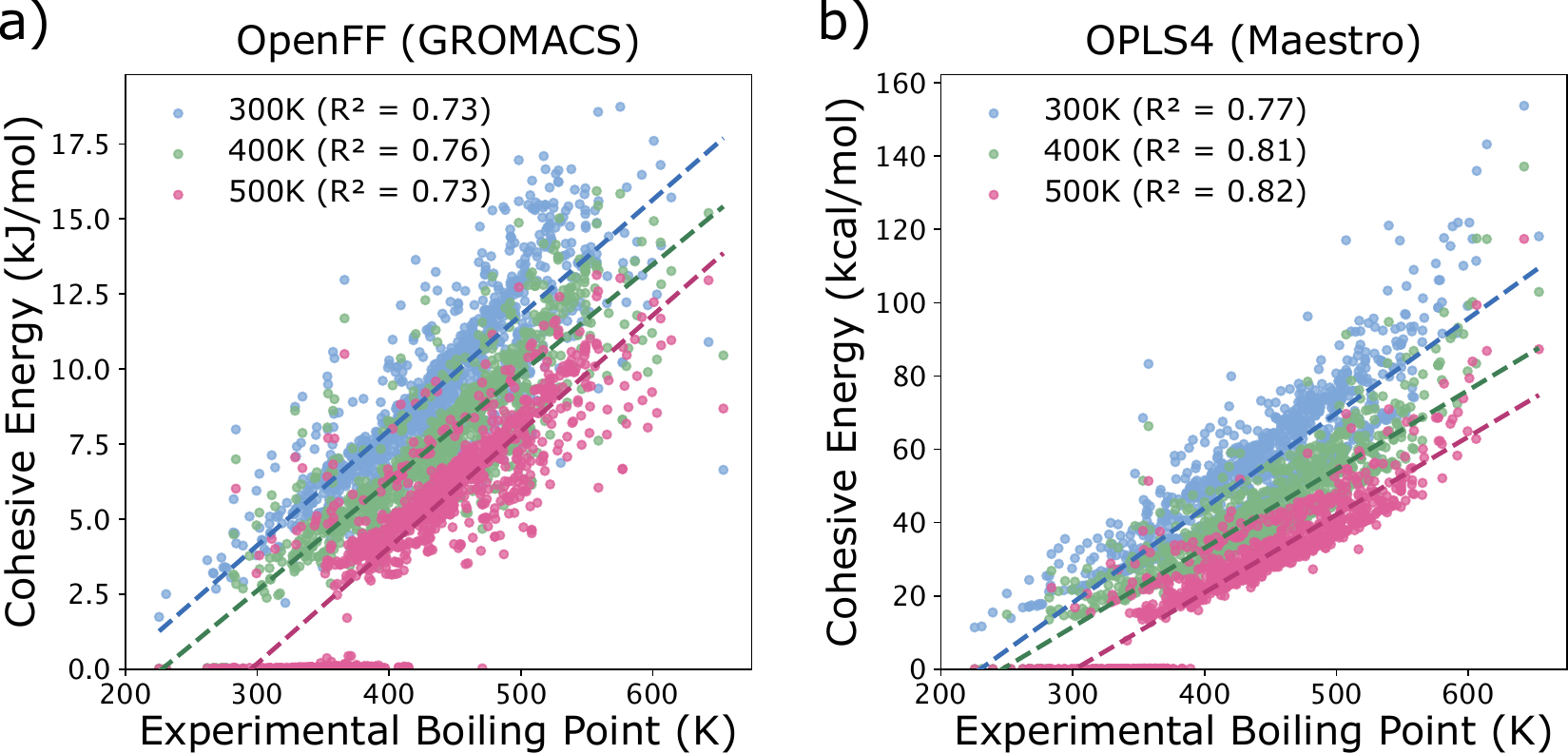}
	\caption{Correlation between simulation-derived cohesive energy ($E_{\text{coh}}$) and experimental boiling point ($T_b$) for 1,280 organic compounds simulated at three temperatures using two independent force fields. (a) OpenFF-2.0.0~\cite{wagner2021openforcefield} results showing linear relationships at 300 K (blue), 400 K (green), and 500 K (pink) with corresponding $R^2$ values of 0.73, 0.76, and 0.73, respectively. (b) OPLS4~\cite{lu2021opls4} results at the same temperature conditions yielding $R^2$ values of 0.77, 0.81, and 0.82. Each data point represents ensemble-averaged intermolecular interaction energies extracted from 20~ns NPT simulations. Data points at near-zero cohesive energy (predominantly at 500K) correspond to compounds that have undergone phase transition to the gas phase during simulation, exhibiting negligible intermolecular interactions due to substantial box expansion and reduced density. They are kept intentionally for the ML model to learn when MD features are meaningful. Linear regression lines are fitted to the full dataset at each temperature, encompassing both liquid-phase and transitioned gas-phase systems.}
	\label{fig: Fig2}
\end{figure*}

{The boiling point of a substance is a direct manifestation of its intermolecular cohesive forces, a foundational principle of classical thermodynamics~\cite{atkins2018physical}. Based on this principle, we hypothesized that descriptors derived directly from the physics of the liquid phase could serve as a more robust foundation for property prediction than purely structural representations. Our approach therefore leverages cohesive energy, extracted directly from atomistic simulations, to encode the intermolecular interaction landscape that governs phase transitions.}

{To validate this hypothesis, our initial objective was to establish a direct, quantitative relationship between the computed cohesive energy and experimental boiling points. We perform atomistic molecular dynamics simulations of 1,280 organic compounds (see section \ref{sec: Materials and Methods}.B) at three temperatures (300, 400, and 500K). To interrogate the robustness of this physics-based descriptor, we compare the performance of two structurally distinct force fields: OpenFF-2.0.0~\cite{wagner2021openforcefield}, a modern open-source parameterization, and OPLS4~\cite{lu2021opls4}, an extensively validated commercial force field, yielding a total of 7,680 independent trajectories (1,280 compounds $\times$ 3 temperatures $\times$ 2 force fields). As shown in Fig.\ref{fig: Fig2}, our analysis reveals a strong linear correlation between the simulated cohesive energy and experimental boiling points, though its quality is force-field dependent~\cite{suarez2024comparative}. This linear trend is physically grounded in Trouton's rule, which implies a direct proportionality between the heat of vaporization ($\Delta H_{\text{vap}}$)---for which our cohesive energy is a proxy---and the boiling temperature ($T_b$)~\cite{atkins2018physical}. Under the OpenFF-2.0.0 force field (Fig.\ref{fig: Fig2}a), linear regression yields coefficients of determination ($R^2$) of 0.73 (300 K; blue), 0.76 (400 K; green), and 0.73 (500 K; pink). This correlation is further enhanced with the OPLS4 force field (Fig.\ref{fig: Fig2}b), which affords $R^2$ values of 0.77 (300 K), 0.81 (400 K), and 0.82 (500 K). More important than the performance difference is the fact that a strong correlation exists for both, especially considering that neither force field was parameterized using boiling points from this dataset. This demonstrates the robustness of cohesive energy as a physics-based descriptor; its utility is not an artifact of a single, specific parameterization but stems from the underlying physical principle it represents. Taken together, this extensive linear analysis confirms that our physics-based approach robustly captures the fundamental intermolecular interactions governing boiling behavior. However, the residual scatter indicates that a simple linear model is insufficient. This is physically expected, as a linear model implicitly assumes a constant entropy of vaporization ($\Delta S_{\text{vap}}$) across all compounds (the basis of Trouton's rule). The observed deviation from linearity thus arises from molecule-specific variations in this entropic contribution. This is precisely where a flexible, non-linear machine-learning model becomes essential, as it can learn the complex mapping from the enthalpic features to the final boiling point while implicitly accounting for these entropic variations.}

{A visible feature in both panels of Fig.~\ref{fig: Fig2} is the emergence of data points with cohesive energies approaching zero, most prominently at the higher simulation temperatures (400 K in green and 500 K in pink). These cases correspond to compounds that undergo vaporization during the NPT simulations, transitioning from a condensed liquid to a dilute gas phase; in this regime, molecules are separated by large distances, intermolecular interactions are negligible, and the computed cohesive energy tends toward zero. Thus, these systems exhibit sharply reduced densities and expanded simulation-box volumes characteristic of the gas state. Importantly, we deliberately retain these phase-transitioned trajectories in the training dataset: they provide explicit supervision linking thermodynamic descriptors---low $E_{\text{coh}}$, low density, high molar volume---to lower boiling points, and delineate the boundary where simulation-derived features remain physically meaningful. In practice, this exposure will enable the CatBoost~\cite{prokhorenkova2018catboost} regressor to implicitly encode liquid–gas phase behavior, enhancing predictive capability for compounds spanning a broad range of volatilities.}

\subsection{Machine learning models trained on simulation-derived descriptors achieve competitive accuracy with substantially reduced feature dimensionality}\label{subsec: Result2}
\begin{figure*}[htb!]
    \centering
	\includegraphics[width=\textwidth]{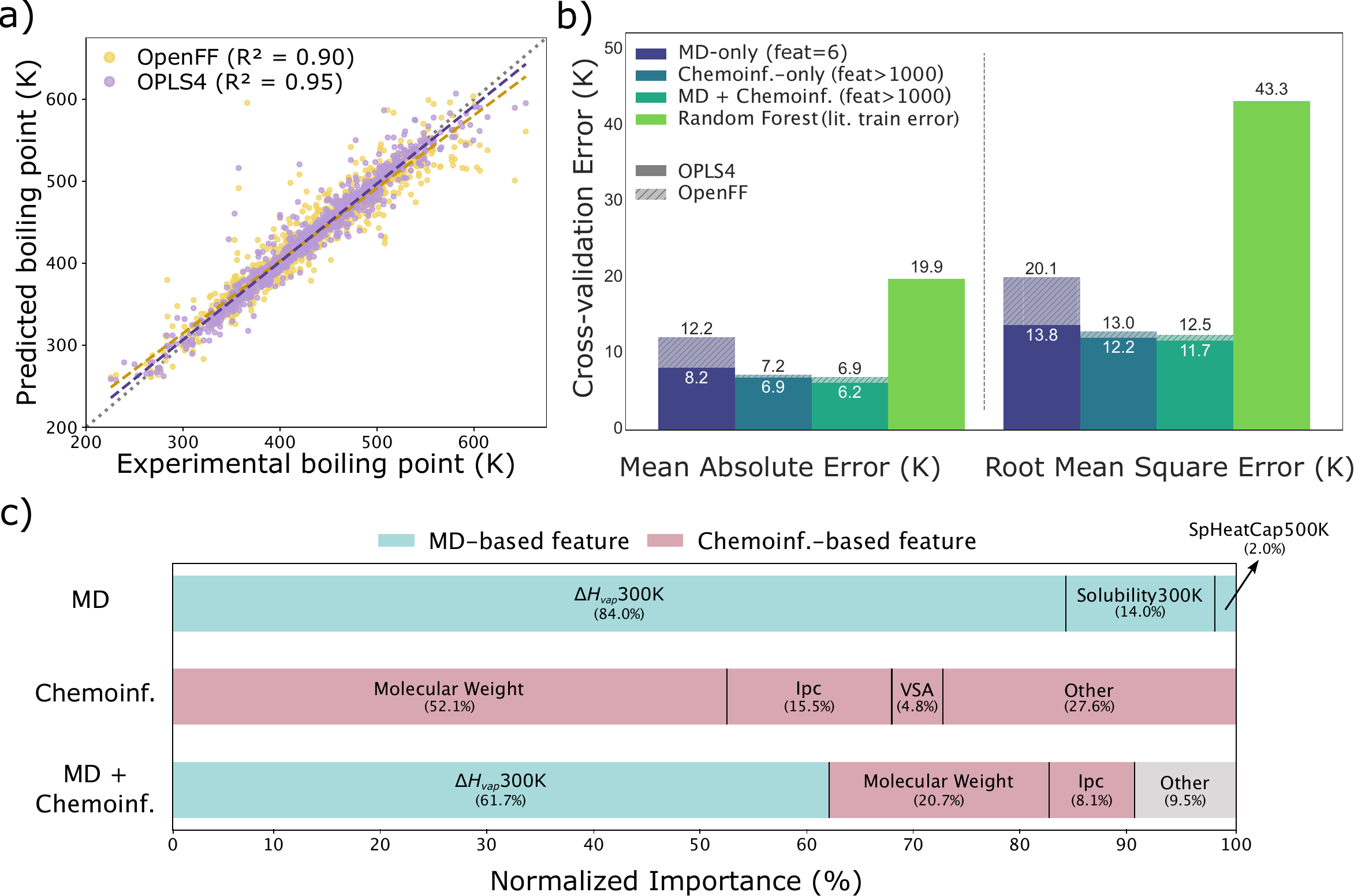}
	\caption{Performance comparison of machine learning models trained on different descriptor sets for boiling point prediction. (a) Cross-validated predicted versus experimental boiling points for the MD-only models. The plot shows the out-of-sample predictions on the 1,280-compound training set, obtained from the 4-fold cross-validation procedure described in the methods. The models were trained exclusively on thermodynamic descriptors from simulations with the OPLS4 (purple) and OpenFF-2.0.0 (yellow) force fields. The solid black line represents a perfect prediction. (b) Cross-validated prediction errors across multiple model architectures. Bars represent mean absolute error (MAE) and root mean squared error (RMSE) for: MD-only models (dark blue), chemoinformatics-only models trained on molecular fingerprints and 2D descriptors (teal), hybrid models combining both descriptor types (dark green), and a literature baseline using Random Forest on the original uncurated dataset (light green)~\cite{kim2024integrating}. Solid bars correspond to OPLS4-derived features, while hatched bars show results for OpenFF-2.0.0 features. (c) Normalized feature importance (in \%) for the OPLS4-based model architectures. The horizontal bars show the relative contribution of the most significant features. Blue segments represent MD-derived features red segments represent chemoinformatics descriptors, and gray represents the cumulative importance of other minor structural features. Top bar (MD-only, 3 features selected): Importance is concentrated in thermodynamic features like the heat of vaporization ($\Delta H_{\text{vap}}$). Middle bar (Chemoinformatics-only, $>$1000 features used): Importance is led by molecular weight, followed by the information content of the characteristic polynomial (Ipc) and van der Waals surface area (VSA). Bottom bar (Hybrid, $>$1000 features used): A synergistic model where a thermodynamic features are complemented by key structural descriptors.}
	\label{fig: Fig3}
\end{figure*}

{Having verified that simulation-derived cohesive energies correlate strongly with experimental boiling points (Fig.\ref{fig: Fig2}), we next evaluate whether these thermodynamic descriptors can serve as effective features for machine learning prediction models. To this end, we trained three categories of CatBoost~\cite{prokhorenkova2018catboost} gradient-boosted regression models (see Materials and Methods S.\ref{sec: Materials and Methods}.D for complete training details): (i) MD-only models using exclusively simulation-derived thermodynamic properties, (ii) chemoinformatics-only models based on molecular fingerprints and 2D structural descriptors, and (iii) hybrid models combining both descriptor types.}

{Figure~\ref{fig: Fig3}A presents the prediction performance of the MD-only models. This parity plot demonstrates that a model trained exclusively on a minimal set of thermodynamic features can yield remarkable predictive accuracy. The model trained on OPLS4-derived features (purple) achieves a strong coefficient of determination of $R^2$ = 0.95, closely tracking the line of perfect prediction. The model based on OpenFF-2.0.0 descriptors (yellow) also performs well, with an $R^2$ = 0.90, though the increased scatter underscores the sensitivity of the approach to force field parameterization, simulation engine and the specific thermostat and barostat algorithms employed in each workflow. The significant improvement in predictive power, moving from the linear correlations in (Fig.\ref{fig: Fig2}a and b) to the machine learning predictions in (Fig.\ref{fig: Fig3}), stems from the model's ability to interpret complex, non-linear thermodynamic behavior. The points with near-zero cohesive energy that appeared as outliers relative to the linear trend in (Fig.\ref{fig: Fig2}) are no longer outliers here. The CatBoost regressor, unlike a simple linear model, successfully learns to associate the full thermodynamic signature of this gas-phase state with a low boiling point. This capability is a primary driver of the model's high predictive accuracy.}

{To contextualize the performance of our MD-only model, we conducted a systematic benchmark against the other architectures, presented in Figure~\ref{fig: Fig3}B. The analysis reveals a clear performance hierarchy. Using the high-fidelity OPLS4 force field, the "MD+Chemoinf." hybrid model, which combines both MD and chemoinformatics features, attains the highest accuracy with a mean absolute error (MAE) of 6.2K and a root mean squared error (RMSE) of 11.7K (dark green bar). It is closely followed by the "chemoinformatics-only" model, trained on a high-dimensional feature set of over 2,000 descriptors, which achieves an MAE of 6.9K and an RMSE of 12.2K (teal bar). Strikingly, our MD-only model (dark blue bar) remains highly competitive with an MAE of 8.2K and an RMSE of 13.8K. This result is highly significant. By replacing thousands of abstract structural features with a handful of physically meaningful descriptors, we reduce the feature dimensionality by over two orders of magnitude. This dramatic reduction is a standard strategy to mitigating the risk of overfitting, as the model is encouraged to learn the governing physical principles rather than memorizing spurious correlations from the training data \cite{dietterich1995overfitting,bejani2021systematic}. This, in turn, is fundamental to its enhanced extrapolative power. Remarkably, this robust generalization is achieved with only a minor sacrifice of 2K in MAE relative to the best-performing, high-dimensional model. The force field dependence is also instructive. The performance of the MD-only model is highly sensitive to the physical accuracy of the simulation, with the MAE degrading from 8.2K with OPLS4 to 12.2K with OpenFF-2.0.0 (hatched dark blue bar). In contrast, the hybrid model shows remarkable resilience, with its MAE only slightly increasing from 6.2K (OPLS4) to 6.9K (OpenFF). This suggests that structural descriptors can partially compensate for inaccuracies in simulation-derived properties, providing a practical pathway for deployment. For a final point of comparison, we include results from the Random Forest model reported by Kim et al.~\cite{kim2024integrating}, the study from which our initial, uncurated dataset was sourced (light green bar) (see section S.\ref{sec: Materials and Methods}.A for more details about the data). This baseline model exhibits substantially higher errors (MAE = 19.9K, RMSE = 43.3K) than even our worst-performing variant. Several factors likely contribute to this performance gap. We hypothesize that two primary factors are: (i) our rigorous data curation produced a cleaner, more consistent training set, and (ii) our physics-based descriptors provide more causally relevant information than purely structural features. However, it must be acknowledged that the observed improvement cannot be attributed solely to the descriptors. The overall modeling strategy itself introduces confounding variables. These include not only the change in the learning algorithm (from Random Forest to CatBoost), but also the distinct feature selection and hyperparameter optimization processes applied to each model. A full deconvolution of these effects is beyond the scope of this study.}

{A deep dive into the feature importance, visualized in Figure~\ref{fig: Fig3}C, reveals the distinct predictive logic of each model and highlights the efficiency of the physics-informed approach. Here, MD-derived features are colored in blue and chemoinformatics descriptors in red. For the MD-only model (top bar), we initially provided a pool of 15 physics-based features: five thermodynamic properties---cohesive energy ($E_{\text{coh}}$), heat of vaporization ($\Delta H_{\text{vap}}$), density ($\rho$), solubility parameter ($\delta$), and isobaric specific heat capacity ($C_P$)---each calculated at three different temperatures (see section S.\ref{sec: Materials and Methods}.B for further details). From this set, CatBoost’s internal feature selection algorithm identifies the most predictive, non-redundant subset. For the model trained on the OPLS4 data, we observe a highly compact feature selection, with the model identifying just three features as sufficient for optimal prediction. As shown in the top bar, the predictive power is overwhelmingly concentrated in the heat of vaporization at 300K ($\Delta H_{\text{vap}}$300K), which accounts for 84\% of the model's cumulative feature importance. The solubility parameter ($\delta$) at 300K (Solubility300K, 14\%) and specific heat capacity at 500K (SpHeatCap500K, 2\%) play minor, supporting roles. This specific three-feature selection, however, highlights the model's data-driven nature rather than a fixed physical rule. The algorithm's goal is to find the most predictive signal, and the composition of that signal is force-field dependent. For instance, when the same modeling process is applied to the features derived from the OpenFF simulations (see Supplementary Information, Fig.S1 in Supplementary Information), the model selects a larger and more varied set of six features. This is instructive, suggesting that for the OpenFF-derived data, the predictive signal is more distributed across several thermodynamic properties, prompting the algorithm to construct a more complex model to capture it effectively.}

{In contrast, the chemoinformatics-based models leverage over 1000 features, but their importance distribution is revealing. In the chemoinformatics-only model, importance is led by molecular weight (52.1\%), but a substantial portion (27.6\%) is fragmented across a vast "long tail" of abstract structural descriptors. A stark contrast emerges in the number of employed features: our model relies on a handful of physical properties, whereas the baseline requires over a thousand abstract ones. The hybrid model, however, reveals the decisive influence of the physics-augmented features. In the OPLS4-based model, a single thermodynamic descriptor---the heat of vaporization at 300K ($\Delta H_{\text{vap}}$)---emerges as the overwhelmingly dominant feature, accounting for 61.7\% of the total importance. This single physical property is nearly three times as important as the next most significant feature, molecular weight (20.7\%), proving it provides the core predictive signal, which is then refined by structural information. This dominance of MD-derived features as the most significant predictive block is consistent across both force fields (see Supplementary Information, Fig.S1).}

{Together, these results establish a powerful principle: the computational investment in generating simulation-derived thermodynamic properties pays significant dividends in predictive power and interpretability. The ability of the MD-only model to distill the prediction down to a few physically intuitive features like $\Delta H_{\text{vap}}$, marks a conceptual advance. This approach transcends the 'black-box' correlation of abstract features by establishing a direct link between the target property and its underlying physical causes. This pursuit of interpretability and causal reasoning is a central goal of modern scientific machine learning~\cite{adadi2018peeking,lavecchia2025explainable}. This work, therefore, not only delivers a high-performance predictive tool but also provides strong support for the development of machine learning models for chemistry that are anchored in first principles based inputs to enhance robustness.}

\subsection{MD-based models demonstrate superior extrapolation to structurally novel and complex chemical systems}\label{subsec: Result3}
\begin{figure*}[htb!] \centering \includegraphics[width=\textwidth]{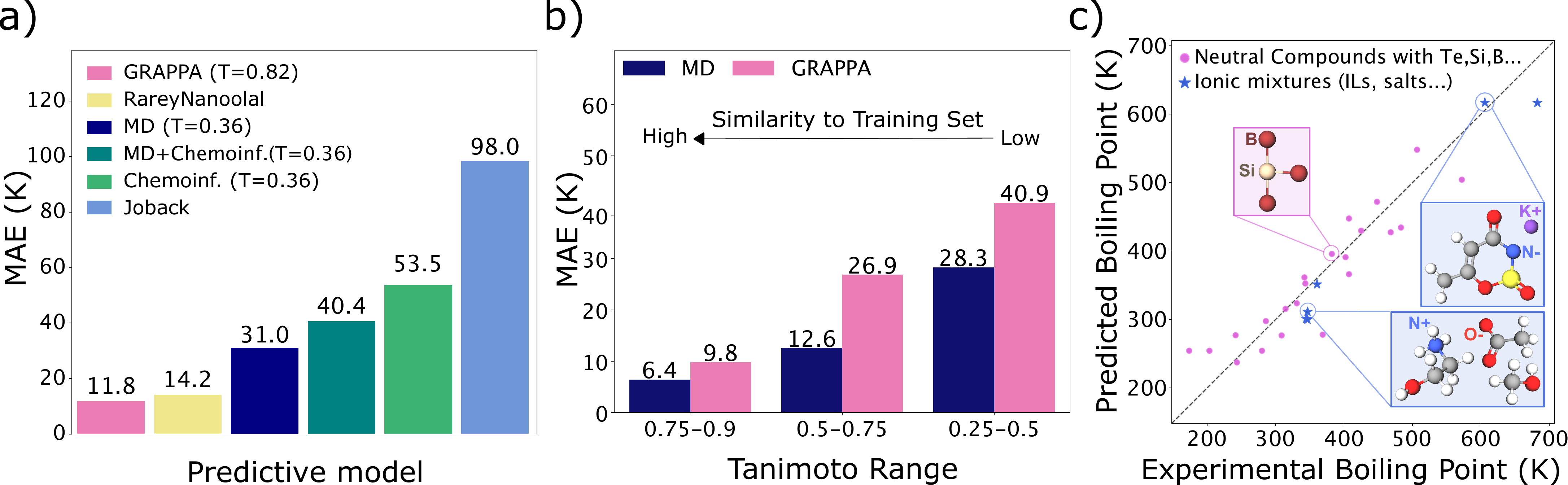} \caption{Extrapolative performance of MD-based models on structurally novel and complex chemical systems. (a) Mean Absolute Error (MAE) on a curated test set of 32 complex organic molecules for various models: GRAPPA (pink), Rarey-Nanoolal (yellow), MD-only (dark blue), hybrid MD+chemoinformatics (dark green), chemoinformatics-only (light green), and the Joback method (light blue). The average Tanimoto similarity of the test set to the respective training sets is noted for GRAPPA (0.82) and our models (0.38), highlighting the greater novelty of the test set for our approach. (b) MAE comparison between the MD-only model (dark blue) and GRAPPA (pink), stratified by Tanimoto similarity between test compounds and each model's respective training set. The MD-only model demonstrates consistently lower error, with its advantage growing significantly as structural similarity decreases. Each bin contains at least 3 data points. (c) Predicted versus experimental boiling points for chemical systems outside the applicability domain of most existing predictors. The model successfully predicts boiling points for neutral compounds with uncommon elements like Si, B, and Te (pink circles, e.g., tribromosilane) and charged systems including salts and ionic liquids (blue stars, e.g., Acesulfame K and an IL-methanol mixture).} \label{fig: Fig4} \end{figure*}

{Given the superior performance of the models trained on OPLS4-derived features, as demonstrated in the previous section, we selected these specific models for the subsequent extrapolation and benchmarking tests. A critical measure of a machine learning model's utility is not just its accuracy within its training domain, but its ability to generalize to new, structurally distinct chemical matter. This capability is of crucial relevance for industrial applications in the pharmaceutical, crop science, and chemical sectors, where the exploration of novel molecular structures is essential for generating intellectual property and discovering materials with unprecedented properties. To evaluate the extrapolative power of our approach, we benchmarked our models against several well-established boiling point predictors: the group-contribution-based Joback~\cite{joback1987estimation} and Rarey-Nannoolal~\cite{nannoolal2004estimation} methods, as well as a state-of-the-art graph neural network, GRAPPA~\cite{hoffmann2025grappa}. A key factor in this selection was their practical accessibility via public web tools that provide instantaneous predictions directly from a SMILES string, such as those for Joback~\cite{joback_calculator} and GRAPPA~\cite{hoffmann2025mlprop}, allowing for a reproducible benchmark. The benchmark was performed on our rigorous external benchmark set (see section S.\ref{sec: Materials and Methods}.A). This set was deliberately curated to probe the limits of extrapolation and consists of 32 structurally diverse active pharmaceutical ingredients with varied complexity sourced from the DrugBank database~\cite{wishart2024drugbank}. These molecules were chosen to represent the kind of structurally complex, real-world chemical matter encountered in industrial discovery pipelines. Crucially, while complex, they are composed of standard organic elements, placing them squarely within the theoretical applicability domain of the chemoinformatic predictors we benchmark. This design ensures a direct and fair head-to-head comparison of true extrapolative performance, with the data also being curated to ensure high-quality experimental values and compatibility across all models. It is important to note the practical trade-off between predictive speed and extrapolative power. While chemoinformatic models are instantaneous, our physics-augmented approach requires an upfront simulation step. However, this computational cost is modest and highly accessible. On a modern GPU, a 20 ns simulation for a single compound is typically completed in half an hour. This makes it feasible to generate descriptors for dozens of compounds simultaneously on a modern multi-core workstation or a small high-performance computing cluster, representing a practical and manageable investment to unlock the model's robust extrapolative capabilities.}

{Figure~\ref{fig: Fig4}A plots the Mean Absolute Error (MAE) in Kelvin for all models on this challenging extrapolation test set. The corresponding Root Mean Squared Error (RMSE) plot, which reinforces the same trends, is provided in the Supplementary Information (Fig.S2.A). A surface-level comparison of the MAE suggests that the literature models GRAPPA and Rarey-Nannoolal achieve the highest accuracy on this dataset.} 

{However, this simple comparison is misleading. A deeper analysis reveals a lower MAE for our thermodynamics-based models once the nature of the extrapolation task is considered. The test set compounds exhibit a high average Tanimoto similarity of 0.82 to GRAPPA's training set, but a low similarity (0.38) to our own. This confirms that for our models, trained on only ~1,000 compounds versus GRAPPA's 20,000, this benchmark represents a significantly more challenging out-of-distribution extrapolation task. In this true extrapolative regime, our MD-only model clearly outperforms the other models trained in this work, achieving an MAE of 31.0 K~\ref{fig: Fig4}A. This result highlights a known limitation of chemoinformatics descriptors: features that are effective for interpolating within a known chemical space become unreliable when the model is faced with novel structures containing fragments or motifs absent from the training data~\cite{tropsha2010best}. The performance degradation is systematic: our chemoinformatics-only model fails (MAE = 53.5 K), and the hybrid model (MAE = 40.4 K) is penalized by including these same unreliable structural features. While a state-of-the-art GNN like GRAPPA does not use traditional chemoinformatics descriptors, its performance is also tightly coupled to training data similarity. Therefore, while GRAPPA remains the superior tool for high-precision predictions within its vast, well-defined domain, our results demonstrate that for true out-of-distribution extrapolation—especially when training data is limited—a model anchored in underlying physics is fundamentally more robust. This finding aligns with recent conclusions in other domains, such as protein-ligand binding affinity prediction~\cite{kubincova2026bridging}.}

{To dissect the true generalization capabilities of both our MD-only model and the GNN we stratified the external test set by Tanimoto similarity, measuring how structurally similar each test compound is to its nearest neighbor in each model's respective training set (a full breakdown is provided in S.I of Supplementary Information). This analysis (Fig.S2.B) reveals their fundamentally different strengths. In the high-similarity regime (Tanimoto $>$ 0.75), GRAPPA's performance is exceptional, with its MAE dropping to 4.1K for compounds most similar to its training data, confirming its power as a high-precision interpolator. To assess true extrapolative capability, we therefore focused the main analysis on the more challenging similarity range below 0.9 (Fig.~\ref{fig: Fig4}B). In this view, a clear and consistent trend emerges where our MD-only model maintains a superior performance. For compounds with moderate similarity (0.75--0.9), our model's MAE is 6.4K versus 9.8K for GRAPPA. This performance gap widens substantially as compound novelty increases, with our model achieving an MAE of 12.6K versus 26.9K in the medium similarity range, and 28.3K versus 40.9K for the most dissimilar compounds.}

{The disparity in robustness is most evident when considering normalized error growth. When extrapolating to the most dissimilar compounds, GRAPPA's error escalates tenfold from its 4.1 K interpolation baseline. In contrast, our MD-only model exhibits a far more controlled response, its error growing by a factor of only 4.4 from its 6.4 K baseline. This confirms their distinct behaviours. While GRAPPA excels at high-precision interpolation, our physics-augmented approach provides a more graceful degradation in performance, establishing it as a more robust foundation for prediction in low-similarity regimes.}

{The ultimate advantage of a physics-based approach is the ability to predict properties for chemical classes that lie entirely outside the domain of traditional models. Existing predictors, including GRAPPA, are restricted to neutral organic molecules containing a limited set of elements (typically C, H, O, N, S, P, and sometimes various halogens) and cannot handle charged species or exotic elements. Fig.~\ref{fig: Fig4}C showcases the model's capability to extrapolate to these systems even though they were not in the training set. The first group (pink circles) includes neutral compounds containing uncommon elements such as Si, B, and Te, or molecules lacking carbon altogether (e.g., tribromosilane). The second group (blue stars) comprises charged systems, including salts (e.g., Acesulfame K), ionic liquids (ILs), and ternary mixtures containing ILs and solvents. We note that experimentally measured boiling points for such systems are scarce, particularly for salts and ILs which often decompose at high temperatures~\cite{ahrenberg2016vapor}. The ability of our model to make reasonable predictions for these complex systems, which are inaccessible to many widely used tools, underscores the power of leveraging fundamental thermodynamic properties for molecular property prediction and opens a path toward a more universal predictive framework.}

{Taken together, this section thus demonstrates the distinct advantages of a thermodynamics-informed framework: it is more robust for extrapolation towards new structures and elemental compositions and enables prediction for chemical domains, such as inorganic and charged species, that lie beyond the reach of conventional methods.}

\section{\label{sec:Conclusion}Conclusions}
{In this work, we have developed a physics-augmented machine learning framework that demonstrates how first-principles-informed networks can perform more robustly when extrapolating. While existing methods like the group contribution-based Rarey-Nannoolal~\cite{nannoolal2004estimation} or machine learning models like GRAPPA~\cite{hoffmann2025grappa} are very competitive around the training set, our work highlights a specific strength in navigating structural novelty. Our approach successfully integrates the regression power of modern machine learning with the fundamental principles of thermodynamics and explicit MD simulations, offering a reliable tool for exploring the novel chemical spaces.}

{We demonstrated that a minimal set of three simulation-derived features yields a model with predictive accuracy competitive with complex chemoinformatics models that rely on thousands of abstract descriptors (Fig.~\ref{fig: Fig3}B). The resulting framework is, moreover, interpretable and physically grounded. This approach successfully integrates the regression power of modern machine learning with the fundamental principles of thermodynamics, offering a reliable tool for exploring novel chemical spaces.}


{The primary strength of our framework is revealed when venturing beyond chemical space of the training set. Our physics-informed model's performance is substantially less sensitive to structural dissimilarity, in contrast to purely data-driven approaches (Fig.~\ref{fig: Fig4}A-B). This robustness establishes our model as a reliable tool, capable of predicting boiling points for systems inaccessible to most established methods, including inorganic compounds, molecules with uncommon elements like silicon and boron, and charged systems such as ionic liquids (Fig.~\ref{fig: Fig4}C). By anchoring predictions to molecular thermodynamics, we demonstrate a clear path toward more generalizable ML models capable of navigating uncharted chemical space.}

{While this study is focused solely on normal boiling point predictions, our results provide strong, systematic evidence for a broader principle: MD-informed machine learning models are more robust for predicting properties governed by intermolecular forces. Rooting predictions in thermodynamics offers a pathway to develop more accurate and widely applicable models, essential for navigating the unexplored chemical territories vital to the future of materials science, pharmacology, drug discovery and engineering.}

\section{Acknowledgements}
N.H.E. acknowledges funding from Bayer Aktiengesellschaft at Kaiser-Wilhelm-Allee 1, 51373 Leverkusen, Germany (“Bayer”). J.R.E. acknowledges funding from Emmanuel College, the University of Cambridge, the Ramon y Cajal fellowship (RYC2021-030937-I), the Spanish scientific plan and committee for research reference PID2022-136919NA-C33, and the European Research Council (ERC) under the European Union’s Horizon Europe research and innovation program (grant agree- ment no. 101160499). J.R. acknowledges funding from the Spanish Ministry of Economy and Competitivity (PID2022-136919NB-C32). This work has been performed using resources provided by the HPC system in Bayer. The authors also thankfully acknowledge RES computational resources provided by Mare Nostrum 5 through the activity 2024-3-0001. Finally, the authors thank Marco Hoffmann for generously providing GRAPPA's training set data, which enabled a direct and valuable comparison between models.
 
\section{\label{sec:Data Availability}Data Availability} To ensure the reproducibility of our research, all relevant data, simulation parameters, and analysis scripts have been deposited in a \href{https://github.com/nurihespejo/BoilingPoints.git}{public GitHub repository}. The repository contains the three datasets used in this work, as detailed in the Materials and Methods section. Furthermore, it provides the GROMACS input files (.mdp) for the energy minimization and production steps of the molecular dynamics simulations based on the OpenFF toolkit. 

\section{References}
\bibliography{library} 

\clearpage

\setcounter{equation}{0}
\setcounter{table}{0}
\setcounter{section}{0}
\setcounter{figure}{0}
\renewcommand\thesection{S\Roman{section}}   
\renewcommand\thefigure{S\arabic{figure}} 
\renewcommand\thetable{S\arabic{table}} 
\renewcommand\theequation{S\arabic{equation}}    
\onecolumngrid
\setcounter{page}{1}
\begin{center}
    {\large \textbf{Thermodynamic Descriptors from Molecular Dynamics as Features for Extrapolable Property Prediction. (Supplementary Material} \\ \par} \vspace{0.3cm}
    Nuria H. Espejo$^{1,2,+}$, Pablo Llombart$^{1,+}$, Andres González de Castilla$^{3}$, Jorge Ramirez$^{4,*}$, Jorge R. Espinosa$^{1,5,6,7*}$, Adiran Garaizar$^{2,*}$ \\ \vspace{0.15cm}
    
    $[1]$ Department of Physical-Chemistry, Universidad Complutense de Madrid, Av. Complutense s/n, Madrid 28040, Spain \\
    $[2]$ Data Science, Bayer AG, Alfred-Nobel-Straße 50, 40789 Monheim am Rhein, Germany \\ 
    $[3]$ Bayer Crop Protection Innovation, Kaiser-Wilhelm-Alle 1, 51373 Leverkusen, Germany \\
    $[4]$ Department of Chemical Engineering, Universidad Polit\'ecnica de Madrid, C/ Jose Gutierrez Abascal 2, Madrid 28006, Spain \\
    $[5]$ Yusuf Hamied Department of Chemistry, University of Cambridge, Lensfield Road, Cambridge CB2 1EW, UK \\
    $[6]$ PhAsIca Biosciences S.L, Calle Velazquez, 27, 28001 Madrid, Spain \\
    $[7]$ Multidisciplinary Institute, Complutense University of Madrid, Paseo Juan XXIII, 1, Madrid 28040, Spain.\\

* = To whom correspondence should be sent.
email: jorge.ramirez@upm.es, jorgerene@ucm.es, adiran.garaizarsuarez@bayer.com 
\end{center}
\thispagestyle{empty}

\date{\today}

\maketitle

\renewcommand\thesection{S\Roman{section}}
\renewcommand\thefigure{S\arabic{figure}}
\renewcommand\thetable{S\arabic{table}}
\renewcommand\theequation{S\arabic{equation}}


\footnotetext{+~Both authors contributed equally.}

\section{\label{sec:s1}Tanimoto Similarity Analysis Methodology} To quantify the structural novelty of the benchmark set relative to different training domains, we performed a Tanimoto similarity analysis using the RDKit toolkit. Morgan fingerprints (256-bit, radius 2) were generated for all molecules in our internal training set, the GRAPPA training set, and our external benchmark set. For each compound in the benchmark set, we calculated its maximum Tanimoto similarity to any molecule in a given training set (either our own or GRAPPA's). This score, ranging from 0 to 1, represents the compound's proximity to the most similar entry in the training domain. This process allowed us to stratify the benchmark set by structural similarity and directly assess how each model's predictive accuracy degrades as a function of increasing chemical novelty, thereby providing a rigorous measure of extrapolative performance.

\section{\label{sec:s3}Supplementary figures}

\begin{figure}[htb!]
    \centering
	\includegraphics[width=.80\linewidth]{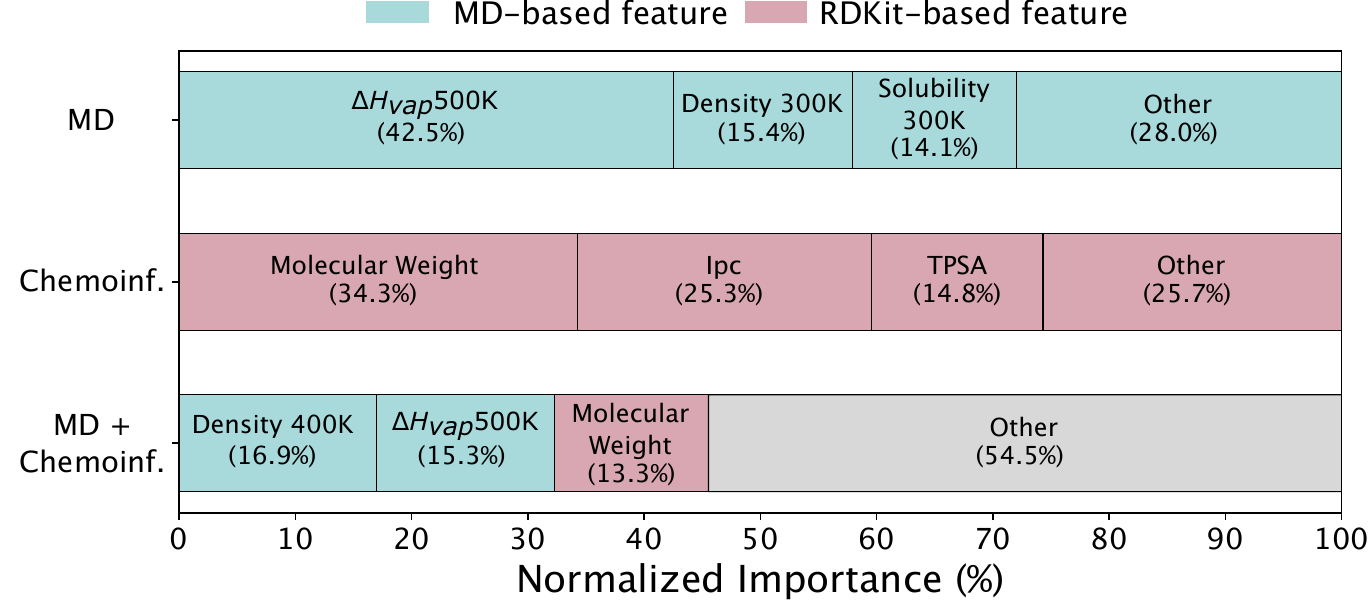}
	\caption{Normalized feature importance (in \%) for the OpenFF-based model architectures. The horizontal bars show the relative contribution of the most significant features. Blue segments represent MD-derived features red segments represent chemoinformatics descriptors, and gray represents the cumulative importance of either minor structural features. Top bar (MD-only, 3 features selected): Importance is concentrated in thermodynamic features like the heat of vaporization ($\Delta H_{\text{vap}}$). Middle bar (Chemoinformatics-only, $>$1000 features used): Importance is led by molecular weight, followed by the information content of the characteristic polynomial (Ipc) and Topological Polar Surface Area (TPSA). Bottom bar (Hybrid, $>$1000 features used): A synergistic model where a thermodynamic features are complemented by key structural descriptors.}  
	\label{fig: FigS1}
\end{figure}

\begin{figure}[htb!]
    \centering
	\includegraphics[width=.80\linewidth]{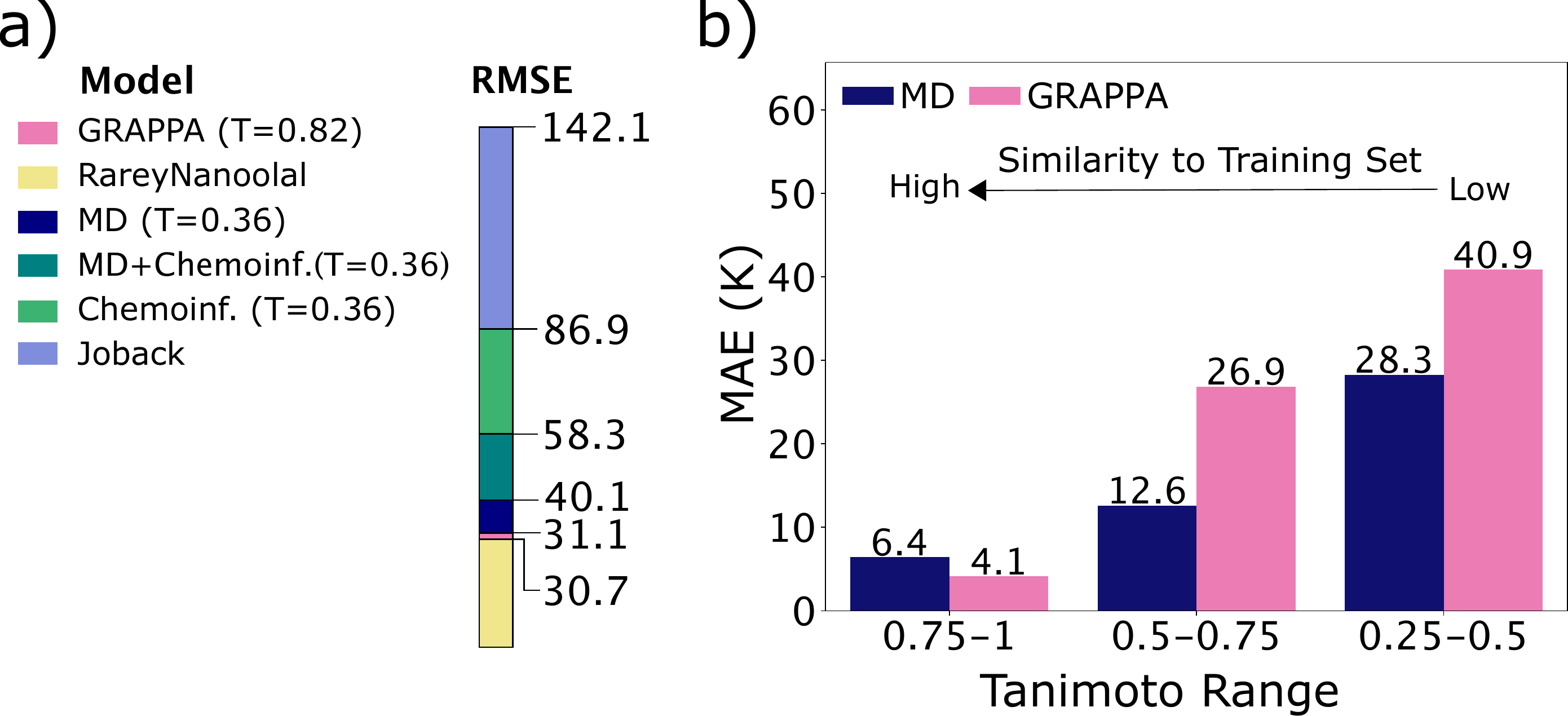}
	\caption{Detailed performance analysis including the high-similarity regime. (a) Root Mean Square Error (RMSE) for the curated test set across various models: GRAPPA (pink), Rarey-Nanoolal (yellow), MD-only (dark blue), hybrid MD+chemoinformatics (dark green), chemoinformatics-only (light green), and the Joback method (light blue). (b) MAE comparison between the MD-only model (dark blue) and GRAPPA (pink), stratified by Tanimoto similarity. This analysis includes the full similarity range up to 1.0, combining the moderate and high similarity compounds into a single bin (0.75-1.0).}  
	\label{fig: FigS2}
\end{figure}

\end{document}